\documentclass[square,aip,preprint,showkeys,groupedaddress]{revtex4}
\usepackage{graphicx}% Include figure files
\begin{document}

\title{Poole-Frenkel effect and Variable-Range Hopping conduction in metal / YBCO resistive switching devices}

\author{A. Schulman}
\thanks{University of Buenos Aires and Conicet scholarships}
\affiliation{Laboratorio de Bajas Temperaturas - Departamento de
F\'{\i}sica - FCEyN - Universidad de Buenos Aires and IFIBA -
CONICET,  Pabell\'on I, Ciudad Universitaria, C1428EHA Buenos Aires,
Argentina}
\author{L. F. Lanosa}
\address{Laboratorio de Bajas
Temperaturas - Departamento de F\'{\i}sica - FCEyN - Universidad de
Buenos Aires and IFIBA - CONICET,  Pabell\'on I, Ciudad
Universitaria, C1428EHA Buenos Aires, Argentina}
\author{C. Acha}
\thanks{corresponding author (acha@df.uba.ar)}
\address{Laboratorio de Bajas
Temperaturas - Departamento de F\'{\i}sica - FCEyN - Universidad de
Buenos Aires and IFIBA - CONICET,  Pabell\'on I, Ciudad
Universitaria, C1428EHA Buenos Aires, Argentina}

\date{\today}

%\draft

\begin{abstract}
Current-voltage (IV) characteristics and the temperature dependence
of the contact resistance [$R(T)$] of Au /
YBa$_2$Cu$_3$O$_{7-\delta}$ (optimally doped YBCO) interfaces have
been studied at different resistance states. These states were
produced by resistive switching after accumulating cyclic electrical
pulses of increasing number and voltage amplitude. The IV
characteristics and the $R(T)$ dependence of the different states
are consistent with a Poole-Frenkel (P-F) emission mechanism with
trapping-energy levels $E_t$ in the 0.06-0.11 eV range. $E_t$
remains constant up to a number-of-pulses-dependent critical voltage
and increases linearly with further increasing the voltage amplitude
of the pulses. The observation of a P-F mechanism reveals the
existence of an oxygen-depleted layer of YBCO near the interface. A
simple electrical transport scenario is discussed, where the degree
of disorder, the trap energy level and the temperature range
determine an electrical conduction dominated by non-linear effects,
either in a P-F emission or in a variable-range hopping regime.

\end{abstract}



\maketitle

\section{INTRODUCTION}

There is a technological need to develop faster, smaller, cheaper
and more reliable memory devices, which is therefore promoting the
search to improve their capacity to retain a higher density of
information, at lower costs, using less energy and operating under
hostile environment conditions.~\cite{Burr08,Yang13} As possible
candidates to fulfil such an overwhelming task, we can find memories
(RRAM) based on the resistive switching (RS) mechanism, where the
non-volatile and reversible resistance state of a metal-oxide
interface is varied upon the application of electric
pulses.~\cite{Sawa08} In the past few years some remarkable
improvements have been made to understand the physics behind the
memory properties of RRAM devices.~\cite{Waser07} As an example, the
underlying memory mechanism of devices based on metal-perovskite
oxide junctions was associated with the resistance change due to
oxygen-vacancy electromigration near the
interface.~\cite{Rozenberg10} Despite these advances, many aspects
are still unclear and should be addressed in order to improve their
practical capabilities. In that sense, one important feature to
determine in each particular device is the conduction mechanism that
dominates its electrical transport properties. This knowledge can be
the key to control most of the desired properties of a memory device
as it points out which are the relevant microscopic factors that
determine their resistive state. Depending on material
characteristics of the metal-oxide interface, the mechanism can be
electrode-limited or bulk-limited.~\cite{Sze06,Chiu14} In the
electrode-limited ones, the work function of the metal, the carrier
affinity and the thickness of the oxide determine the barrier height
and the probability to produce an electric-field-induced-current
through the junction. In this case, the mechanism can be described
as Schottky, Fowler-Nordheim or direct tunneling emission. In the
bulk-limited case, the conduction mechanism is determined by the
electrical properties of the oxide near the interface, particularly
by the existence of traps and their energy levels. Poole-Frenkel
emission (P-F) and space-charge-limited conduction (SCLC) are two
examples of transport mechanisms influenced by the energy
distribution and density of traps.

Metal-YBa$_2$Cu$_3$O$_{7-\delta}$ (YBCO) interfaces on ceramic and
thin-films samples have both shown bipolar RS
characteristics.~\cite{Acha09a,Acha09b,Placenik10,Acha11} Their
retentivity~\cite{Schulman11,Placenik12} as well as their response
to cyclic electric field stresses~\cite{Schulman12} have been
previously studied. Although the microscopic origin of its RS
properties was successfully associated to the electromigration of
oxygen vacancies~\cite{Rozenberg10}, no detailed studies of the
conduction mechanism through the interface have been already
performed.

In this paper, our goal is to identify the relevant transport
mechanism of Au / YBa$_2$Cu$_3$O$_{7-\delta}$ (YBCO) interfaces, in
order to point out the microscopic factors that determine the
resistive state, as the temperature dependence and magnitude of the
resistance change upon producing a RS. Our results indicate that a
P-F emission mechanism dominates the current-voltage dependence of
the junction, consequently indicating the existence of carrier traps
and of a low conductivity region in the interfacial YBCO. Within
this scenario, most probably related to a random distribution of
oxygen vacancies near the junction, we present a description of the
electrical conduction based on a modified Variable-Range Hopping
(VRH) mechanism. The modification that we propose is based on
considering that the available electrical carriers are only those
thermally or voltage assisted de-trapped which obey to a P-F law. In
this framework, we show that increasing the amplitude or the number
of applied pulses has qualitatively the same microscopic effect on
modifying the trap energy level as well as the resistivity and the
geometric factor of a conducting channel.

\section{EXPERIMENTAL DETAILS}

The device was prepared by sputtering four metallic electrodes on
top of a face of a good quality and optimally-doped YBCO textured
ceramic slab (8$\times$4$\times$0.5 mm$^3$, see the inset of
Fig.~\ref{fig:protocolo}). The YBCO slab was synthesized following a
similar procedure to the one described in Ref.~\cite{Porcar97} The
homogeneity and the oxygen content was checked by measuring its
resistive superconducting transition, which showed a 10-90\% width
of $\simeq$ 2 K and a $T_c^{onset} =$ (90 $\pm$ 0.2) K (see
ref.~\cite{Acha09a}). By considering the relation between $T_c$ and
the mean oxygen content (7-$\delta$)\cite{Takagi87,Cava90}, we
estimated that $\delta \lesssim$ 0.05. The sputtered electrodes have
a 1x1 mm$^2$ area and a mean separation of 0.5 mm. Copper leads were
carefully fixed by using silver paint, without contacting directly
the surface of the sample. Au and Pt were chosen as metals for the
pair of pulsed electrodes, labeled $1$ and $2$, respectively. As we
have previously shown~\cite{Schulman13}, the Pt / YBCO interfaces
have a lower resistance value than the Au / YBCO ones
($R_{Pt}\lesssim R_{Au}/3$), and a small RS amplitude. Thus, we may
disregard the influence of the Pt / YBCO electrode and proceed as if
only the Au / YBCO electrode were active (i.e. presents a relevant
RS effect), simplifying the effects produced upon voltage pulsing
treatments.

In order to analyze the sensitivity of the microscopic parameters
that control the electrical transport properties of the Au / YBCO
interface to the voltage amplitude and to the number of applied
pulses, we applied the following pulsing protocol (see
Fig.~\ref{fig:protocolo}):

%fig1
\begin{figure}
\vspace{0mm}
\centerline{\includegraphics[angle=0,scale=0.5]{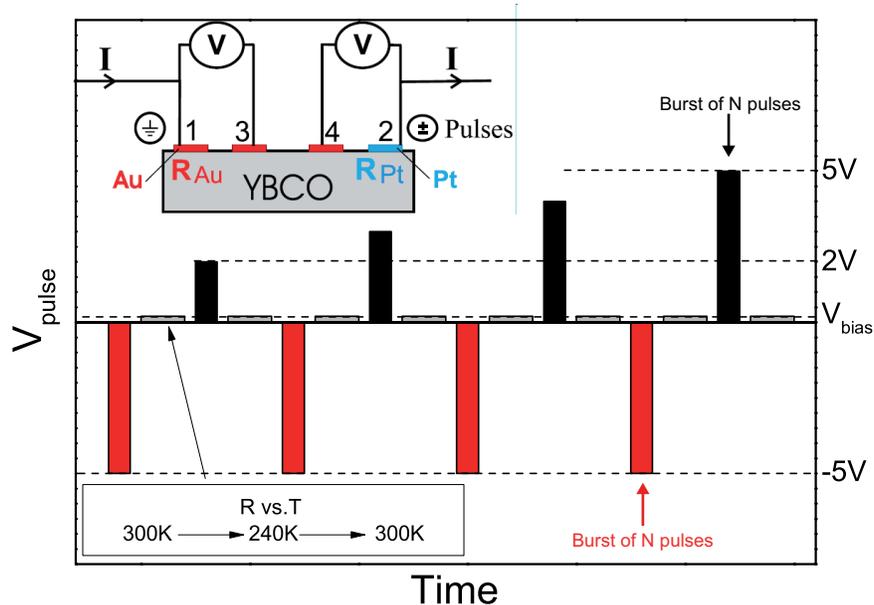}}
%\centerline{\includegraphics[width=6.7cm,height=5.7cm]{R4wycontactos4.eps}}
\vspace{-5mm}\caption{(Color online) Diagram of the pulsing protocol
used to study the effect of the amplitude and of the number of
pulses on the RS of the Au / YBCO interface. The black bars
correspond to the reset process while the red ones to the set. Each
bar represents a burst of N pulses. The grey areas correspond the
$R(T)$ measurement with a small bias current; the protocol was
repeated for each number of pulses per burst (N). The inset shows
the YBCO ceramic device and its electrode configuration. Au was
sputtered on pads 1-3-4 while Pt on pad 2. See the text for the
details of the measuring procedure.} \vspace{-0mm}
\label{fig:protocolo}
\end{figure}

Temperature is initially stabilized at 300 K, with the active Au /
YBCO electrode set to its low resistance state ($R_{Au} = R^{LRS}$).
A Pt thermometer is used to determine the temperature of the device.
To produce a RS to a high resistance state ($R_{Au} = R^{HRS}$), a
"reset" burst of $N$ positive unipolar pulses (500 $\mu$s width at 1
kHz rate) of a $V_{pulse}$ amplitude ($2 V \leq V_{pulse} \leq 5 V$)
is applied with an Agilent 33250A waveform generator to electrodes
1-2, during a time $t_0$ (from 10 s to 500 s, depending on the $N$
value). We want to note here that the polarity of the pulses was
defined arbitrarily with the ground terminal connected to the active
Au / YBCO contact. With this definition, for example, a negative
pulse produces a positive voltage drop on the Au / YBCO interface.
In order to avoid overheating effects on the resistance
measurements, a time equal to $t_0$ is waited before measuring the
temperature dependence of remnant resistance $R^{HRS}(T)$. Then this
resistance is measured as a function of temperature, cooling down to
240 K and heating back to 300 K at a 2 K per minute rate. For that,
a DC three terminal method is used, by applying a small bias current
(100 $\mu$A) to electrodes 1-2 and by measuring the voltage
difference ($\Delta V <$ 0.1 V) between electrodes 1-3. A Keithley
224 was used as current source and an Agilent 34420A as
nanovoltmeter. Voltage was also measured between electrodes 2-4 to
check the small RS effect of the Pt / YBCO electrode.

Also using this configuration, DC current-voltage characteristics
(IV) were measured at 300 K, where a small I-V range was explored in
order to avoid a RS of the active electrode. Corrections to
$R^{HRS}$ by considering the resistance of the bulk YBCO (beyond the
interfacial zone) are negligible taking into account that its value
is only $\simeq 0.1 \Omega \ll 50 \Omega < R_{Au}$ in this
temperature range.

After the $R^{HRS}(T)$ measurement, due to the bipolar nature of the
RS in the metal/perovskite interface, a "set" burst of maximum
opposite polarity (-$V_{pulse}^{max} = -5 V$) of the same number of
pulses of the "reset" burst is applied to come back to a LRS and to
erase the cumulative effects of the reset pulses. Then, the
temperature dependence of the LRS remnant resistance is measured
[$R^{LRS}(T)$] following the same procedure as with $R^{HRS}(T)$.
The process is then completely repeated for a new $V_{pulse}$ value,
increased with a fixed step, until it reaches our experimental
maximum ($V_{pulse}^{max} = 5 V$). The whole process was performed
for different numbers of pulses $N$ in the $10^4 \leq N \leq
5~\times~10^5$ range.

\section{RESULTS AND DISCUSSION}

\begin{figure}
\vspace{7mm}
\centerline{\includegraphics[angle=0,scale=0.4]{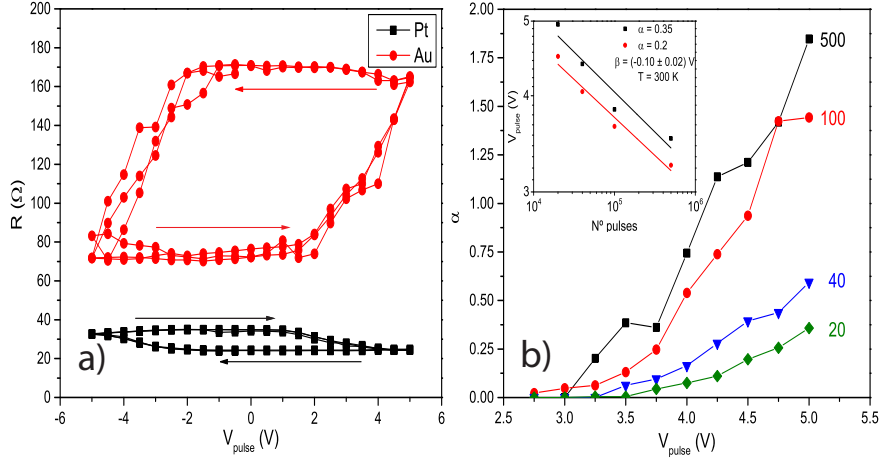}}
%\centerline{\includegraphics[width=6.7cm,height=5.7cm]{R4wycontactos4.eps}}
\vspace{-5mm}\caption{(Color online) a) RHSL for both pulsed
contacts (N = 100~$\times~10^{3}$). The Pt / YBCO interface shows
lower resistance and smaller RS amplitude than Au / YBCO.  b)
Relative variation of the remnant resistance [$\alpha =
(R^{HRS}-R^{LRS})/R^{LRS}$] of the Au / YBCO interface as a function
of the amplitude of the reset pulses ($V_{pulse}$) with different
number of square pulses per burst: $N =$ 20~$\times~10^{3}$,
40~$\times~10^{3}$, 100~$\times~10^{3}$, 500~$\times~10^{3}$. The
inset shows the analog to the Basquin curves ($V-N$ curves) at 300K
for two arbitrarily failure criteria $\alpha =$ 0.2 and 0.35 (see
ref.~\cite{Schulman13}).} \vspace{-0mm} \label{fig:alpha}
\end{figure}

In Fig.~\ref{fig:alpha} a) the hysteretical remnant resistance of
both Au / YBCO and Pt / YBCO contacts ($R_{Au}$ and $R_{Pt}$,
respectively) after applying a burst of 100~$\times~10^{3}$ pulses
of amplitude $V_{pulse}$ can be observed. As $V_{pulse}$ is varied
following a loop, these curves are usually called Resistance
Hysteresis Switching Loop (RHSL). The observed bipolar memory
behavior of both contacts and their complementary response to
$V_{pulse}$ is a typical characteristic of the RRAM
devices.~\cite{Acha09b,Rozenberg10,Acha11} It can be noticed the
lower switching amplitude and the better stability of $R_{Pt}$ in
comparison to $R_{Au}$, in accordance to the expected passivity of
the former. Taking this into account, hereafter our results and
discussions will be only related to the Au / YBCO active electrode.

In Fig.~\ref{fig:alpha} b), the relative amplitude of the remnant
resistance change at room temperature, defined as
$\alpha=(R^{HRS}-R^{LRS})/R^{LRS}$, is presented as function of
$V_{pulse}$ and for different numbers of applied pulses ($N$). As
shown in a previous study~\cite{Schulman13}, $\alpha$ increases
logarithmically with increasing $N$ and follows a power-law-like
dependence with $V_{pulse}$ (if $V_{pulse}$ is higher than a
$N$-dependent threshold voltage, $V_c(N)$). A similar result has
been observed for Ag / manganite interfaces and was interpreted as a
consequence of the voltage and history dependent spatial
distribution of oxygen-vacancies near the interface.~\cite{Ghenzi12}

%fig2

The inset of Fig.~\ref{fig:alpha} shows the $V_{pulse}$-N curves
that produce different $\alpha$ values (0.2 and 0.35) which recalls
a similar behavior to that described by the Basquin law of
mechanical stress-lifetime test, which corresponds
to~\cite{Schutz96,Suresh98} $V_p \sim N^\beta$, with $\beta \simeq$
-0.1.

\begin{figure}
\vspace{7mm}
\centerline{\includegraphics[angle=0,scale=0.9]{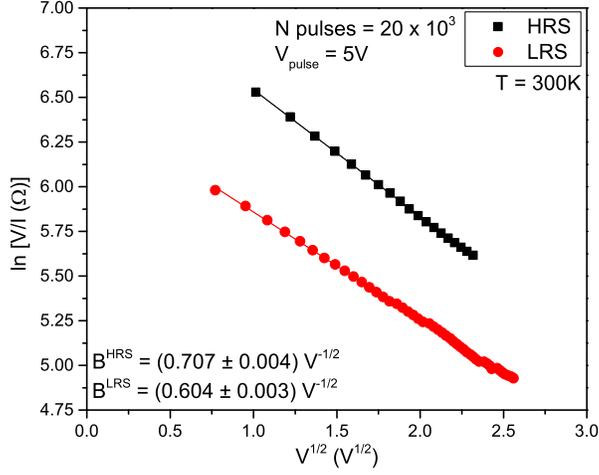}}
%\centerline{\includegraphics[width=6.7cm,height=5.7cm]{R4wycontactos4.eps}}
\vspace{-5mm}\caption{(Color online) ln ($V/I$) plotted against
$V^{1/2}$ for the LRS and the HRS, achieved with a set and a reset
burst of $N = 20~\times10^{3}$ pulses with  $V_{pulse} = \pm 5 V$ at
300 K, respectively. The lines are fits to the P-F emission
(Eq.~\ref{eq:P-F2}). The fitted $B$ values of each state are
indicated.} \vspace{-0mm} \label{fig:RvsV}
\end{figure}

From the IV characteristics obtained for both states at 300 K, we
have plotted a typical result as $ln(V/I)$ vs $V^{1/2}$ in
Fig.~\ref{fig:RvsV}. Data is well fitted by a linear dependence
although there are some small deviations at low voltages. This
result indicates that bulk-limited P-F emission is the dominant
mechanism for the electrical transport of carriers through this
interface. This can be seen by considering the expression of the P-F
current ($I_{P-F}$) as a function of the voltage $V$ at a fixed
temperature $T$, which corresponds to~\cite{Simmons67}

\begin{equation}
\label{eq:P-F} I_{P-F} = (R_0)^{-1} V \exp[-E_t/(k_B T) + B
V^{1/2}],
\end{equation}
\noindent with
\begin{equation}
B = \frac{q^{3/2}}{k_B T (\pi \epsilon d)^{1/2}},
\end{equation}

\noindent where $R_0$ is a pre-factor that will be discussed later,
associated with the geometric factor of the conducting path, the
electronic drift mobility ($\mu$) and the density of states in the
conduction band. $E_t$ is the trap energy level, $k_B$ the Boltzmann
constant, $q$ the electron charge, $\epsilon$ the dielectric
constant of the oxide and $d$ the distance between electrodes. The
voltage-dependent contact resistance can then be expressed as:

\begin{equation}
\label{eq:P-F2} R = \frac{V}{I_{P-F}} = R_0 \exp[E_t/(k_B T) - B
V^{1/2}].
\end{equation}

From the $B$ values obtained by fitting the data presented in
Fig.~\ref{fig:RvsV} with Eq.~\ref{eq:P-F2} and by assuming that the
dielectric constant of YBCO~\cite{Mannhart96} is of the order of
$\epsilon \simeq$ 200$\times$$10^{-12}$ F/m, the characteristic
distance $d$, where most of the voltage drops occurs, can be
estimated. We obtain that $d^{HRS}$ $\simeq$ 0.8 $\mu$m and $d^{LRS}
\simeq$ 1.1 $\mu$m, which corresponds to a size much lower that the
distance between contacts ($\sim$ 500 $\mu$m), in accordance to
previous results that indicate that the RS-active-region remains
limited to a small zone near the interface.~\cite{Wang12}

%fig3

As the $R(T)$ measurements were performed using low currents, the
remnant non-linear (NL) contact resistance (Eq.~\ref{eq:P-F2})
should be nearly ohmic when $V\ll V_L$, where $V_L = \pi \epsilon d
E_t^2/q^3$. In that case, Eq.~\ref{eq:P-F2} can be simplified as

\begin{equation}
\label{eq:Rrem} R \simeq R_0 \exp[E_t/(k_B T)].
\end{equation}

\begin{figure}
\vspace{0mm}
\centerline{\includegraphics[angle=0,scale=0.9]{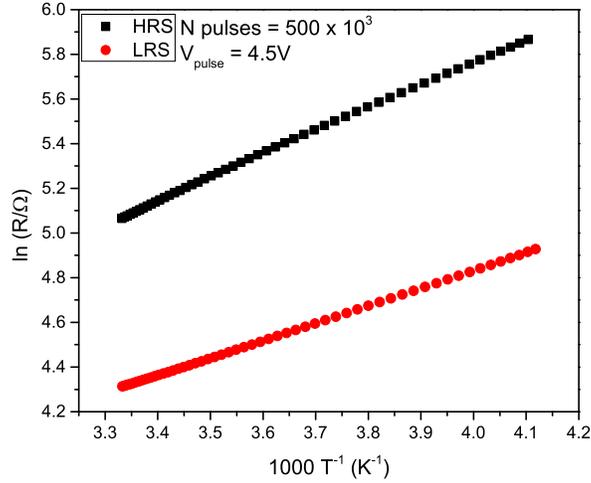}}
%\centerline{\includegraphics[width=6.7cm,height=5.7cm]{R4wycontactos4.eps}}
\vspace{-5mm}\caption{(Color online) $ln (R)$ vs. $T^{-1}$ for the
LRS and the HRS, achieved at room temperature with a set and a reset
burst of $N = 500~\times~10^{3}$ pulses with  $V_{pulse} = \pm 5 V$,
respectively.} \vspace{-0mm} \label{fig:RvsT}
\end{figure}

\begin{figure}
\vspace{-5mm}
\centerline{\includegraphics[angle=0,scale=0.9]{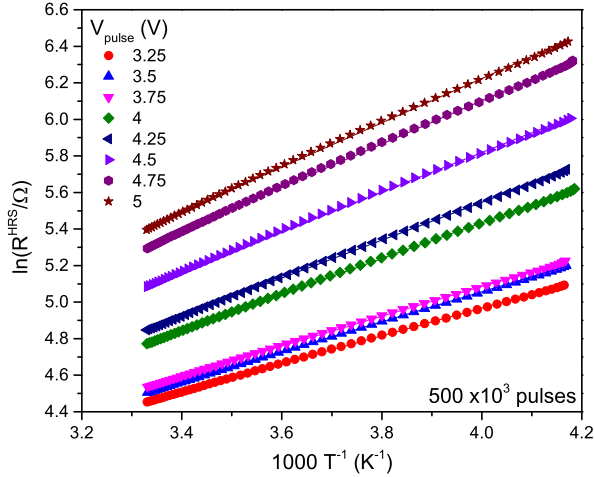}}
%\centerline{\includegraphics[width=6.7cm,height=5.7cm]{R4wycontactos4.eps}}
\vspace{-0mm}\caption{(Color online) $ln (R)$ vs. $T^{-1}$ for the
HRS, achieved at 300 K with a reset burst of $N = 500~\times~10^{3}$
pulses with varying pulse amplitudes.} \vspace{-0mm}
\label{fig:RvsT2}
\end{figure}

Fig.~\ref{fig:RvsT} shows the temperature dependence of the Au /
YBCO contact resistance in both states (LRS and HRS) plotted as
ln(R)vs $T^{-1}$ in the 240 K to 300 K temperature range. The
$V_{pulse}$ sensitivity of $R^{HRS}(T)$ can be observed in
Fig.~\ref{fig:RvsT2}, where ln(R)vs $T^{-1}$ is shown for varying
$V_{pulse}$ and a fixed $N = 500~\times~10^{3}$. Similar results
were obtained for the whole set of $N$ values explored (not shown
here). The nearly straight line followed by the data in this
temperature range indicates a good agreement with Eq.~\ref{eq:Rrem},
implying that $R_0$ and $E_t$ should be practically
temperature-independent. $E_t(V_{pulse})$ and $R_0(V_{pulse})$ for
$N = 500~\times~10^{3}$ pulses can then be extracted for both states
(LRS and HRS) by using Eq.~\ref{eq:Rrem} to fit the data presented
in Fig.~\ref{fig:RvsT} and Fig.~\ref{fig:RvsT2}. A similar procedure
was followed for other values of $N$.

%fig4

%fig5

\begin{figure}
\vspace{0mm}
\centerline{\includegraphics[angle=0,scale=0.35]{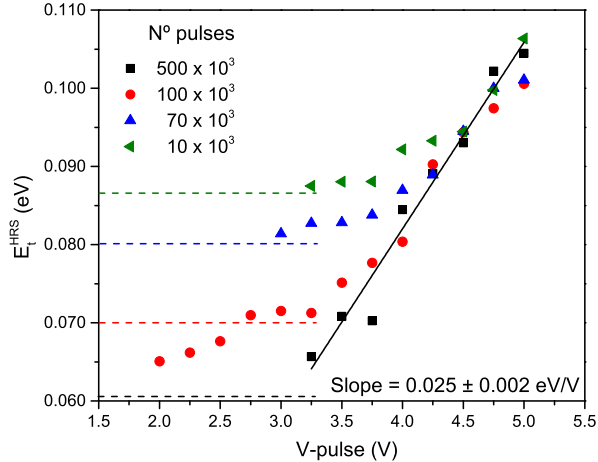}}
%\centerline{\includegraphics[width=6.7cm,height=5.7cm]{R4wycontactos4.eps}}
\vspace{-5mm}\caption{(Color online) Trap energy level of the HRS
($E_t^{HRS}$) as a function of $V_{pulse}$, for different number N
of applied pulses per burst. The dashed lines correspond to the
values of the trap energy level for the LRS ($E_t^{LRS}$), which are
practically $V_{pulse}$-independent. The solid line indicates a
linear fit of the data for $V_{pulse} > V_c(N)$} \vspace{-0mm}
\label{fig:gapvsV}
\end{figure}

The fitting $E_t^{HRS}(V_{pulse})$ values are shown in
Fig.~\ref{fig:gapvsV} for the whole set of $N$ pulses per burst
explored. The obtained $E_t$ values for both states and for all
pulsing treatments are within the 0.06 eV to 0.11 eV range, which
validates the approximation made to obtain Eq.~\ref{eq:Rrem} ($V_L
\sim$ 10 V). It can be observed that in the low voltage region
[$V_{pulse} < V_c(N)$], $E_t^{HRS}$ is nearly voltage independent,
but decreases with increasing $N$ as a consequence of the decrease
of $E_t^{LRS}$, due to the set pulsing treatment (see
Fig.~\ref{fig:protocolo}). For $V_{pulse} \geq V_c(N)$, $E_t^{HRS}$
increases linearly with $V_{pulse}$ with a slope of 0.025 eV
V$^{-1}$ and becomes less $N$-dependent. This voltage dependence
with a relatively low N-dependence is in accordance to the
previously observed behavior~\cite{Schulman13} , mentioned as a
Basquin-like law. This law indicates that the amplitude of the RS
change ($\alpha$) is more sensible to a voltage variation than to a
number of pulses change. In this way, to produce the same RS effect,
$N$ should be increased by 3 orders of magnitude if $V_{pulse}$ is
reduced to half its value.

%fig6

\begin{figure}
\vspace{0mm}
\centerline{\includegraphics[angle=0,scale=0.9]{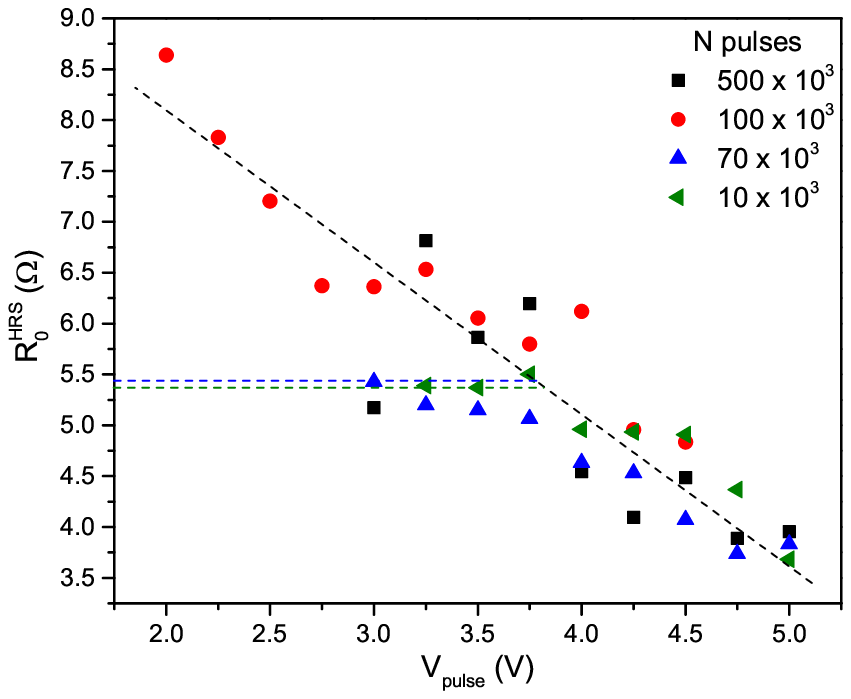}}
%\centerline{\includegraphics[width=6.7cm,height=5.7cm]{R4wycontactos4.eps}}
\vspace{-0mm}\caption{(Color online) $R_{0}$ for the HRS as a
function of $V_{pulse}$ for varying burst lengths. The dashed lines
are the values of $R_{0}$ for the LRS, which are nearly
$V_{pulse}$-independent.} \vspace{-0mm} \label{fig:R0vsV}
\end{figure}

The obtained $R_0(V_{pulse})$ dependence for different bursts
lengths is shown in Fig.~\ref{fig:R0vsV}. Besides the noisy
behavior, two regions can be observed as for $E_t$: a voltage
independent region for $V_{pulse} < V_c(N)$ and, surprisingly, a
tendency to decrease for $V_{pulse} \geq V_c(N)$. \\

%fig7

On one hand, as our results indicate that the main conduction
mechanism through the Au/YBCO interface is a P-F emission, we have
to consider that transport properties are dominated by a
carrier-trap region in the YBCO part of the junction. In that sense,
YBCO in this interface-active-region cannot be in a metallic-like
state, even at room temperature or in the LRS, which indicates that
the average density of oxygen-vacancies ($\delta$) should be roughly
in the 0.7 to 1.0 range.~\cite{Wuyts96} For these $\delta$ values,
YBCO is not superconducting and has resistivities in a broad range,
from $\sim$~4 to $\sim$~4000
m$\Omega$cm.~\cite{Matsushita87,Milliken00} Oxygen vacancy zones can
then be considered as the positively-charged traps that will capture
electrons, impeding hole mobility. They are, up to now, associated
with the microscopic origin of the bipolar RS effect and models
based on their electromigration manage to describe nontrivial
experimental results.~\cite{Rozenberg10} Here, the fact that $E_t$
increases with increasing $V_{pulse}$ may indicate that oxygens are
not removed uniformly from the YBCO lattice, but in a correlated
manner, increasing the size of oxygen-depleted spots, probably as
the local electric field is larger in these zones. As it will be
discussed later, the decrease of $R_0$ with $V_{pulse}$ can be
related to an increase of the cross-section area of a low-conducting
filament in the HRS.

On the other hand, it was shown~\cite{Matsushita87,Milliken00} that
the electrical conduction of low-oxygen-content YBCO is by a 2D
variable range hopping (2D-VRH) mechanism, as oxygen vacancies
introduce disorder in the electronic potential.~\cite{Efros84} At
the same time, we may consider the scenario here depicted, where
part or the totality of the hopping carriers can be trapped in
deeper potential wells, associated with the electric-pulse-induced
oxygen vacancy spots. In this case, their number will be limited,
introducing a characteristic de-trapping energy ($E_t$) and the NL
characteristics of the P-F effect.

The standard 2D-VRH resistivity ($\rho^{VRH}$) of the
oxygen-depleted YBCO near the interface can be expressed as:

\begin{equation}
\label{eq:sigma} \rho^{VRH} = \rho_0 \exp[(T_0/T)^{1/3}],
\end{equation}
\noindent where $T_0$ is a parameter related to the localization
length and to the density of states at the Fermi energy, and
\begin{equation}
\label{eq:sigma0} \rho_0 = (k_B T) / (n q^2 \nu_{ph} R_h^2),
\end{equation}
\noindent where $n$ is the density of carriers, $q$ the electron's
charge, $\nu_{ph}$ a characteristic phonon-assisted hopping
frequency and $R_h$ the hopping distance.

This expression should be modified to consider the influence of
traps. If the density of traps $n_t \geq n$ and if we assume that at
T=0 K all the electrical carriers are trapped, only those de-trapped
or ionized ($n_i$) will participate in the VRH conduction. As the
barrier to overcome ($E_t$) can be reduced by the P-F effect, then
$n_i = n \exp(-E_t / k_BT + BV^{1/2})$. Within this scenario, the
contact resistance associated with an oxygen-depleted YBCO zone of
characteristic length $d$ and a conducting area $S$ can be rewritten
as:

\begin{equation}
\label{eq:Rc} R = \frac{d}{S} \rho^{VRH} \exp[E_t/(k_B T)-BV^{1/2}]
= \frac{d}{S} \rho_0 \exp[(T_0/T)^{1/3}+E_t/(k_B T)-BV^{1/2}].
\end{equation}

%fig8
\begin{figure} [b]
\vspace{0mm}
\centerline{\includegraphics[angle=0,scale=0.48]{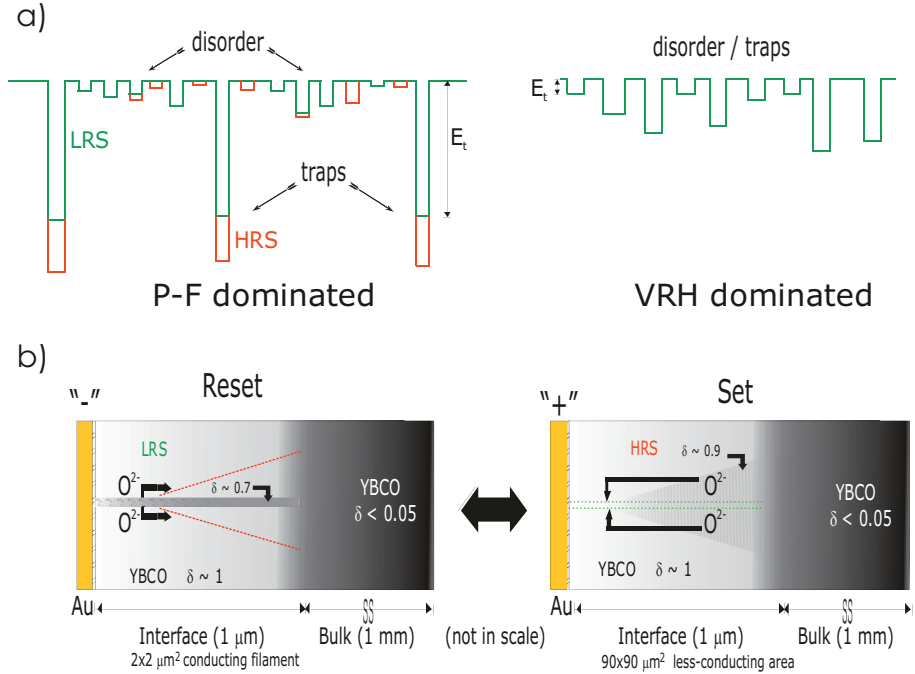}}
%\centerline{\includegraphics[width=6.7cm,height=5.7cm]{R4wycontactos4.eps}}
\vspace{5mm}\caption{(Color online) a) Schematic representation of
the spatial distribution of the disordered electronic potential
background with deep traps (P-F dominated transport) and with
shallow traps (VRH dominated transport). In the former case,
increasing the amplitude of electric pulses produce deeper traps,
clearly differentiated from the background, while in the latter
case, traps and background disorder are indistinguishable, and b)
Sketch of the Au / YBCO interface. The P-F IV characteristics
indicate that the interfacial YBCO corresponds to a $\sim$
1000$\times$1000$\times$1 $\mu$m$^3$ volume with a random
distribution of vacancies, although spots of many surrounding
vacancies acting as carrier-traps should be formed. After the set
pulsing treatment, a $\sim$ 2$\times$2$\times$1 $\mu$m$^3$
conducting filament is formed in the LRS by electromigration of
oxygens from the bulk, in accordance to previous
models.~\cite{Rozenberg10} The oxygen content of this filament can
be increased up to the optimum oxygen content, creating a
superconducting filament of reduced geometric factor.~\cite{Acha09a}
Oxygens from this filament are then removed when applying the reset
pulses, generating the HRS, which corresponds to a less conducting
and thicker filament ($\sim$ 90$\times$90$\times$1 $\mu$m$^3$),
probably of conical shape.~\cite{Kim09} } \vspace{-0mm}
\label{fig:esquema}
\end{figure}

In fact, as the disorder that produces the VRH conduction can be
associated with an oxygen-vacancy random distribution, which may
produce trapping centers for the carriers that yield to a P-F
emission, both conduction mechanisms may coexist or dominate,
depending on different factors: as stated in Eq.~\ref{eq:Rc}, VRH
should be the main mechanism for temperatures $T > T_L$ with $T_L =
(\frac{(E_t/k_B)^3}{T_0})^{1/2}$.  While for $T < T_L$ the
conduction should be of the P-F type. In the same way, NL effects
should be observed for voltages $V \gtrsim V_L$ in the P-F case, and
may even be detected when $V \gtrsim 0.1V_L$. While in the VRH
regime, they should be noticed for a low temperature region, where $
T \lesssim \frac{q^{9/4} V^{3/4}}{k_B^{3/2}(\pi \epsilon d)^{3/4}
T_0^{1/2}}$. On other words, depending on the degree of disorder, on
the trap energy level and on the temperature range explored,
low-oxygen-content YBCO can present, intrinsically, a VRH or a P-F
regime, as well as NL-IV characteristics. The VRH $+$ the NL-IV
regime was in fact observed previously~\cite{Matsushita87} for
YBa$_2$Cu$_3$O$_6$, where the electrical transport properties,
measured in a four terminal configuration at 40 K to 300 K, were
well described by a 2D-VRH conduction and showed the presence of a
NL regime at low temperatures, whose origin has not been discussed
so far. Conversely, our results in the Au / YBCO junctions for the
240 K to 300 K temperature span indicate that the P-F emission is
favored over the VRH regime. In that way, the interfacial region of
YBCO may present higher trap energy levels (oxygen-depleted spots)
and a less disordered crystal structure than that existing in the
oxygen-depleted YBCO with $\delta \sim 1$. As mentioned, this can be
related to a correlated oxygen vacancy distribution, which increases
the trap energy level without increasing the overall disorder, as in
a random vacancy distribution. A sketch of these possible two
scenarios is presented in Fig.~\ref{fig:esquema}a).

The mentioned low conductivity of the pristine YBCO in contact with
the Au interface may be a consequence of the higher oxidation energy
of Au when compared to YBCO. One possibility (see
Fig.~\ref{fig:esquema}b) is that Au depleted oxygens from the
interfacial YBCO, generating a disordered potential with traps for
the charge carriers.

Just to give an order of magnitude, we can assume that this
interfacial YBCO has a mean density of vacancies of $\delta \simeq$
0.7 to 1 per unit cell volume ($n_t\sim$  6$\delta$$\times$10$^{21}$
cm$^{-3}$) in the LRS and in the HRS, respectively. In this case,
$\rho^{VRH}$ should range from 4 m$\Omega$cm to 4000
m$\Omega$cm.~\cite{Wuyts96,Matsushita87} In these conditions, the
conduction area $S$ can be roughly estimated by considering
Eq.~\ref{eq:Rc} and the low voltage measurements of the contact
resistance at 300 K. The obtained values, $S^{LRS} \sim$ 2$\times$2
$\mu$m$^2$ and $S^{HRS} \sim$ 90$\times$90 $\mu$m$^2$, indicate the
filamentary nature of the area modified by the RS pulsing
treatments. The dispersive diffusion of oxygens may be the reason of
the increase of $S^{HRS}$ when compared to $S^{LRS}$. Although
$\rho^{VRH}$ for YBCO increases with increasing $\delta$, the
increase of $S^{HRS}$ determines the decrease of $R_0$ with
$V_{pulse}$ (see Fig.~\ref{fig:R0vsV}). Local measurements of the
oxygen content near the interface are needed to prove the validity
of this sketched scenario.

\section{CONCLUSIONS}

In summary, transport measurements performed on Au / optimally-doped
YBCO as a function of temperature and electric field showed an
electrical conduction dominated by a P-F emission mechanism. As this
is a bulk property of the interfacial YBCO, we inferred that in a 1
$\mu$m region near the interface, YBCO is oxygen-depleted, in a way
that favors the existence of traps for the carriers. The energy
level of these traps ($E_t$) can be increased linearly by increasing
the amplitude of the voltage pulses. This result can be interpreted
as an indication that the electromigration of oxygens may not be
produced randomly but in a correlated manner. The number of applied
pulses $N$ produces less changes on $E_t$ than those obtained by
increasing $V_{pulse}$, in accordance to a previously stated
similarity with the stress-lifetime Basquin law. We also proposed a
simple description of the intrinsic transport scenario for
oxygen-depleted YBCO, where the VRH carriers may be trapped by
oxygen vacancies, yielding to P-F emission or to VRH conduction,
with or without NL effects, depending on the temperature range
explored. Finally, an estimation of the conduction area for both
states reveals the filamentary nature of the zone influenced by the
pulsing treatments.

\section{ACKNOWLEDGEMENTS}
We would like to acknowledge financial support by CONICET Grant PIP
112-200801-00930, PICT 2013-0788 and UBACyT 20020130100036BA
(2014-2017). We also acknowledge V. Bekeris and F. Acha for a
critical reading, and D. Gim\'enez, E. P\'erez Wodtke and D.
Rodr\'{\i}guez Melgarejo for their technical assistance.

%\bibliography{bibRRAM}

\end{document}